# How the global structure of protein interaction networks evolves

**Andreas Wagner**

University of New Mexico
Department of Biology
167A Castetter Hall
Albuquerque, NM 817131-1091
Phone: +505-277-2021
FAX: +505-277-0304
Email: wagnera@unm.edu

**Abstract**

Two processes can influence the evolution of protein interaction networks: addition and elimination of interactions between proteins, and gene duplications increasing the number of proteins and interactions. The rates of these processes can be estimated from available *Saccharomyces cerevisiae* genome data and are sufficiently high to affect network structure on short time scales. For instance, more than 100 interactions may be added to the yeast network every million years, a substantial fraction of which adds previously unconnected proteins to the network. Highly connected proteins show a greater rate of interaction turnover than proteins with few interactions. From these observations one can explain – without natural selection on global network structure – the evolutionary sustenance of the most prominent network feature, the distribution of the frequency $P(d)$ of proteins with $d$ neighbors which is a broad-tailed distribution consistent with a power law ($P(d) \propto d^{-\gamma}$). This distribution is independent of the experimental approach providing information on network structure.

Post-genomic biology is unravelling a wide variety of biological circuitry, ranging from metabolic networks to transcriptional regulation and protein interaction networks (1-7). Once the structure of a genetic network is known, interlaced questions arise about its functions and evolutionary origin. Does the network's structure tell us anything about the network's function? How could natural selection have shaped its global structure? Or does natural selection act largely on smaller, local scales and thus play only a minor role in shaping the network as a whole?

Before addressing any of these questions, one has to represent and characterize a network's structure. Any choice between multiple possible representations is best guided by the nature of the information available. For the best characterized genetic networks this information is purely qualitative – who interacts with whom – lending itself to the simplest possible representation, that of a graph. Graphs are mathematical objects consisting of nodes and edges. In a protein interaction graph, for example, two nodes (proteins) are connected by an edge (they are adjacent) if they interact physically. The *degree* or *connectivity* $d$ of a protein is the number of other proteins it interacts with. A *path* between two proteins $v_0$,, $v_i$ is a sequence of adjacent proteins $v_0, v_1, \ldots, v_{i-1}, v_i$ leading from $v_0$ to $v_i$. The number of edges in this path is called the pathlength. There are many ways to characterize the structure of graphs, including the distribution of path lengths, the number of cyclic paths, and various measures of clumping of nodes into clusters of highly connected nodes (8). The simplest possible measure is that of the number of edges per protein, and its distribution in the graph. For protein interaction networks, as for a variety of other graphs (9), this distribution appears to follow a power law (10, 11). That is, when choosing a protein at random from the network, the probability $P(d)$ that this protein has $d$ interaction partners is proportional to $P(d) \propto d^{-\gamma}$, $\gamma$ being some constant characteristic of the network. The same class of distribution is observed in metabolic network graphs (3-5).

This feature of cellular networks raises questions about its origin and purpose. Does a power law degree distribution convey any kind of advantage to an organism? If so, then natural selection has probably acted on this distribution itself to shape it into a power law. Such a selectionist perspective has been put forth recently (12). It is based on the observation that power law degree distributions can endow a network with robustness against perturbations. Upon removal of randomly chosen nodes from the network, the mean path length in such a network is affected less than when perturbing a network with different degree distributions (12). Other network features are similarly robust (3, 5). For metabolic networks, a possible advantage of small mean path lengths stems from the importance of minimizing transition times between metabolic states in response to environmental changes (13-15). Networks with robustly small average path lengths may adjust more rapidly to environmental perturbations. A key prediction of this selectionist explanation is that the removal of highly connected nodes would affect an organism more severely than that of lowly connected nodes. Data consistent with this prediction exists for the yeast protein interaction network (11). However, this observation is equally consistent with more pedestrian explanations, such as pleiotropic effects of highly connected proteins regardless of network structure. In addition, recent work suggests that highly connected proteins are not subject to more severe evolutionary constraints, as would be expected under this selectionist explanation (16, 17).

**A null-hypothesis about the origin of global network structure**. Ideally one would like to explain the persistence of any organismal feature directly from the evolutionary processes affecting it. To do this for the yeast protein interaction network is the goal of this contribution. How does this network sustain a power law degree distribution, when processes such as mutation and gene duplication constantly erode this distribution? In the face of such perturbations, would natural selection on this distribution not be essential to sustain the degree distribution?

It is often stated that any null-hypothesis explaining an organismal feature must not involve natural selection or any optimality criterion. Natural selection is to be invoked only if all such null-hypotheses are to be rejected. To provide such a null hypothesis, I will first consider the processes that influence the structure of the yeast protein interaction network and estimate their rates from empirical data. Based on this information, I will then explain the degree distribution without invoking natural selection on this distribution.

**Global network features are independent of experimental approach.** The biological interpretation of protein interaction networks as produced by genome-scale interaction screens has been hampered by several factors. First, they collapse spatial and temporal information into one freeze-frame static image of the network. Second and more critically, independent large-throughput experiments with very similar experimental designs generate interaction maps with a limited number of common interactions (1, 2). Despite these shortcomings, protein interaction maps can already be used successfully to predict the spatial expression domains and functional annotations of many proteins from their interaction partners (18). They thus clearly contain biologically useful information.

Although great uncertainty is associated with individual interactions identified by genome-scale experiments, global statistical features of protein interaction networks do not depend on the veracity of each identified interaction, and may thus contain the most reliable information. Figures 1a and 1b show the distribution *P(d)* for the two yeast protein interaction map reported by Uetz and collaborators (2) and Ito and collaborators (1). These networks were generated using the yeast two-hybrid assay (19) but otherwise different experimental designs. Although they show limited overlap in protein interactions, their degree distributions are identical. Importantly, they are both consistent with a power law ($P(d) \propto d^{-\gamma}$) with statistically indistinguishable exponents (Uetz: $\gamma$=2.55±0.35; Ito: $\gamma$=2.43±0.35). Proteins highly connected in one data set are also highly connected in the other (Pearson r=0.52, $P \ll 10^{-3}$, df=329; Spearman $r_s$=0.31, $P \ll 10^{-3}$). In addition, publicly available protein interaction data generated with experimental approaches different from the two-hybrid assay also generate a network with identical power law degree distribution (Fig. 1c; $\gamma$=2.67±0.2). The degree distribution of protein interaction networks is a global structural feature robust to the vagaries of experimental approach.

Two separate processes can influence such global network features, gene duplications generating new proteins, and turnover of interaction between existing proteins. I will now discuss these two processes in turn.

**Network evolution by gene duplication.** Individual gene duplications occur at formidable rates in eukaryotic genomes and are thus potentially important factors in influencing the structure of protein interaction networks (20, 21). In yeast, the rate of gene duplications is approximately equal to 52 duplications per genome and million years ($8.3 \times 10^{-3}$/gene/Myr). As many as 90% of gene duplicates are likely to eventually get lost after duplication (22, 23), leading to an effective duplication rate closer to $8.3 \times 10^{-4}$/gene/Myr. Gene conversion is not rampant in the evolution of these duplicates (24), and the bulk of yeast duplicate gene pairs (paralogues) have low to moderate expression and thus low codon usage bias. For these paralogues, the time elapsed since duplication can be roughly estimated through the accumulated rate $K_s$ of synonymous substitions per synonymous site. In yeast a $K_s$=1 corresponds to approximately 100 Myrs since duplication (10). While any such divergence estimates (especially for $K_s$>1) are imprecise, they are here used only for a coarse grouping of gene pairs, or to eliminate highly divergent pairs. Moreover, all reported results depend only on order-of-magnitude estimates of this and other evolutionary rates.

Are genes whose products have many protein interactions less likely to undergo duplication? If so, genome evolution through gene duplication would be intertwined with the degree structure of the protein interaction network. One could not be understood without the other. However, there appears to be no association between protein degree and the propensity of the respective gene to become duplicated (Fig. 2a).

Fig. 2b shows a protein P with four interactions. When the gene encoding this protein undergoes a duplication, P and its duplicate $P^*$ still have four neighbors each. But the number of interactions of each of their neighbors has now increased by one. A gene duplication always increases – never decreases – the degree of proteins. The proteins whose degree increases are the interaction partners of the duplicated proteins. This simple observation implies that a power law distribution of degree could not be sustained under the influence of gene duplications alone. In a network that has such a degree distribution to begin with, the relative frequency of proteins with one interaction partner (which constitute the bulk of the observed network; Fig. 1) would slowly approach zero. The frequency of proteins with degree 2 would follow, and so forth, leading to proteins of ever-higher degree to dominate the network. But are gene duplications even sufficiently frequent to influence network structure on an evolutionarily relevant time scale? One glance at the abundance of duplicate gene products in the network (Fig. 2c, 47% of genes in the network have paralogues with $K_s$<3) shows that this influence of gene duplications must be profound.

Why, then, are there any network proteins with low degree? Many aspects of a gene's function (25-27) tend to get lost rapidly after gene duplication. Similarly, protein interactions diverge rapidly between such genes. Fig. 2d shows a time-course of this divergence, where paralogous genes are grouped into bins according to their divergence ($K_s$), corresponding to time since duplication. The ordinate shows the fraction $f$ of shared interactions between duplicates. This is the number of interactions two duplicates have in common, divided by the total number of interactions of the two duplicates. Only for the most recent duplicates ($K_s<0.5$) is this number moderately large ($0.5<f<0.6$). For more distant duplicates $f<0.15$, i.e., they share less then 15 percent of interactions (Fig. 2d). Two proteins chosen at random from the network have an expected $f$ of $1.4\times10^{-3}$ ($\sigma=1.2\times10^{-3}$). The figure covers only divergences up to $K_s=1$ but more distantly related proteins show even smaller $f$. The binning interval of $K_s=0.25$ in the figure was made possible by pooling data from all three data sets (1, 2, 28). However, each data set individually also shows this pattern: $f<0.1$ when averaged over gene pairs with $0.5<K_s<3$ in each of the two-hybrid data sets (1, 2) and an average $f$ of 0.159 for the network derived from non-two hybrid data (28).

In sum, most shared interactions between paralogous genes have diverged within 50Myrs after duplication. The prevalence of degenerative mutations after gene duplication suggests that most of this divergence is due to mutational loss of interactions. What is the effect of this divergence on the evolution of the degree distribution? Fig. 3 shows a numerical analysis addressing this question. In this simulation I subjected the protein interaction network as reported in ref. 2 to recurrent duplication of randomly chosen genes. After each duplication, gene duplicates were allowed to lose common interactions as observed in the data. Even after 1000 and 2000 duplications (corresponding to approximately 1.2 Gyrs and 2.4 Gyrs of evolution, respectively), the degree distribution of the network was unaffected. Thus, while gene duplications would have a profound effect on network structure, this effect disappears once subsequent interaction divergence is taken into account. An additional case in point is the observation that members of one gene family are not overrepresented among interaction partners of highly connected proteins, as would be expected if gene duplication substantially influenced network evolution (results not shown). The task of accounting for a persistent power-law distribution thus reduces to explaining this distribution from the second major evolutionary process influencing it.

**Evolution by addition and deletion of interactions.** Addition and elimination of physical interactions between proteins is caused by mutations that change protein surfaces, be it point mutations, insertions or deletions. The products of such mutations may then be retained by natural selection or genetic drift. Here is the most important question about this process: Is the rate of interaction turnover sufficiently high to influence network structure? To be sure, interactions appear to get lost rapidly after gene duplications (Fig. 2d). However, because organisms tolerate many degenerative mutations after gene duplications, such a high rate of interaction loss may not be representative of any "background" rate of interaction turnover independent of gene duplication. Can this rate be estimated? The following is such an estimate, based on the rate at which interactions are added to the network. It relies on the observation that new protein interactions occasionally evolve between the products of paralogous genes.

I will discuss this estimate in detail for one of the reported networks (2), and then summarize the results for the other two (1, 28). In this network, there are 15 paralogous gene pairs ($K_s$<3) whose products interact with each other ("cross-interactors"). A number of these gene pairs also show self-interactions of their members, as might occur if a protein forms homodimers. Figure 4 shows how self and cross-interactions can evolve after gene duplications. First, a gene product may have been a self-interactor before duplication. In this case, observed self-interactions and cross-interactions are remnants of the self-interaction before duplication. Second, a cross-interaction may have evolved *de novo* after the duplication. Which of the observed 15 cross-interactions have formed *de novo*? The left-most column of numbers in Figure 4b shows the number of paralogues in yeast with $K_s$<3 that have the indicated combination of self-interactions and cross-interactions. There are at least two conspicuous features of the data. First, both proteins show self-interaction for only one out of 19 paralogous pairs. The one gene pair with both self- interactions and a cross-interaction is an old duplicate with $K_s$>1, indicating that its three interactions may have evolved in the more than 100 Myrs passed since the duplication. Secondly, for 10 out of 15 paralogous pairs with cross-interactions, neither duplicate shows self-interactions. Thus, there is an abundance of paralogous gene pairs whose features are more easily explained if cross-interactions between duplicates evolve *de novo*. But what if most of the listed duplicates originated from self-interacting proteins, and natural selection has since preferentially eliminated self-interactions? The following suggests otherwise. The propensity to undergo duplication is similar for self-interacting and all other proteins (37% vs. 43% of these proteins have paralogues). The fraction of self-interacting proteins in the network is small, less than 4.5%. If selection eliminated self-interactions preferentially after duplication, then the fraction of paralogues with self-interactions should be even smaller. But this is not the case. To the contrary, 47% of the duplicate proteins shown in Fig. 4b (left column) display at least one self-interaction.

Based on these observations, I will assume that cross-interactions have evolved since the duplication for all but the protein pairs where both partners also show self-interactions. (The more conservative assumption that only pairs with no self-interaction reflect de novo evolution of interactions would somewhat lower the estimates below, but would not affect any of the qualitative conclusions.) There are 14 gene duplicates that fit this bill, among a total of 79 gene duplicates with (i) $K_s<3$, (ii) both duplicates being part of the network, and (iii) each gene family with more than two members represented by only one member pair. Item (iii) eliminates bias due to including paralogues from one gene family more than once. This leads to an estimate of $(14/79)(1/300)=5.9\times10^{-4}$ newly formed interactions per protein pair and per $\Delta K_s=0.01$ ($\approx$1Myr). Extrapolating to the total of $4.99\times10^5$ total protein pairs ($n(n-1)/2$ for $n=999$) in this network, one would expect $(5.9\times10^{-4})(4.99\times10^5)=294.5$ added interactions per Myr. While it can not be said with certainty that this rate is uniform, it is important that the rate does not appear accelerated immediately after duplication: Only 2 of the 14 observed cross-interactions between paralogues involve paralogues with $K_s<0.5$.

The rate at which interactions are formed is similar for the network reported by Ito and collaborators (1). 15 cross-interacting duplicates among 58 paralogues yield an estimated rate of $8.6\times10^{-4}$ new interactions per protein pair per $\Delta K_s=0.01$, or a total of 270 newly formed interactions among the $3.14\times10^5$ possible protein pairs per 1Myr. For the non-two hybrid data (28), there are 12 cross-interacting duplicates among a total of 83, leading to $4.8\times10^{-4}$ newly formed interactions per protein pair per Myr, or a total of 108 newly formed interactions per Myr for the $2.26 \times 10^5$ protein pairs in the network. Again, the observed cross-interactions between paralogues do involve only one and zero paralogues with $K_s<0.5$, respectively, indicating that the rate of interaction gain is not elevated shortly after duplication.

This data, however crude, shows that the rate at which new interactions are added to the network is high, upward of 100 added interactions per Myr. It also has another important implication. Assume that there was a drastic imbalance between the rate, $c_+$, at which new interactions are added and the rate, $c_-$, at which interactions are eliminated. If, say $c_+:c_-=2:1$, a network might sustain a net gain of more than 50 interactions per Myr, leading to a doubling of the number of interactions within 20 Myrs. Conversely, if $c_+:c_-=1:2$, the number of interactions would drop by one half in less than 10 Myrs. Thus, the number of interactions per node would either vanish or explode within an evolutionarily short amount of time. There is no evidence for such drastic change. For example, the protein interaction map for the prokaryote *Helicobactor pylori*, established with a variation of the experimental design generating two of the maps analyzed here, suggests that proteins do not have vastly different numbers of interactions in these two organisms (29, 30). This indicates that the rates of interaction gain and loss must be approximately equal ($c_+ \approx c_-$).

**Power law degree distribution through local rules.** Interaction turnover without gene duplication is sufficiently rapid to influence network structure drastically. The next question is whether it alone can sustain a power law degree distribution. A variety of models have been proposed in which addition and deletion of edges can generate power law degree distributions (reviewed by 9). None of them relies on any global selection principle favoring networks with power law degree distributions over other networks. Put differently, in all studied models power law degree distributions emerge only through local addition and deletion of nodes and edges. Merely two general principles are sufficient to obtain networks with power law degree distributions (9). First, nodes must be added to a network, even if only occasionally. Second, new interactions must be more likely to involve highly connected nodes than nodes with few connections. The latter principle is also referred to as preferential attachment (“the rich get richer”) (31). Are these two features observed for protein interaction networks?

The first question is whether new nodes get occasionally added to the network. When considering both non-two hybrid and two-hybrid data together, one finds 32 edges among non-self interacting paralogues with $K_s<3$, involving 21 proteins that have no other interactions. This is an indication that a substantial fraction of edge additions may add previously unconnected proteins, an observation that also holds for each of the available data sets separately (results not shown).

Second, do new interactions between proteins already in the network preferentially involve highly connected proteins? Fig. 5 relates the degree of a protein to the likelihood that the protein obtains a new interactions. It is again based on cross-interacting paralogues with $K_s<3$. While the data is not sufficient to make precise estimates of a proportionality constant, it shows a strong and nearly linear correlation between the degree of a protein, and the likelihood of acquired new interactions. Preferential attachment does occur in protein interaction networks.

In sum, the key prerequisites – node addition and preferential attachment – to obtain power law degree distributions through local interaction rules are met for protein interaction networks. This permits the following model for network evolution. The model assumes that interactions are added between network proteins (at a rate $c_{+e}$), that interactions are added between network proteins and proteins not in the network (at a rate $c_{+n}$), and that interactions are eliminated from the network (at a rate $c_-$). Consistent with preferential attachment, edges are preferentially added or eliminated with a probability linearly proportional to the degree $d$ of proteins they are attached to. The model does not include gene duplications, because these do not distort the degree distribution (Fig. 3). For networks growing through node addition the expected degree $d(s,t)$ at time $t$ of a protein that has been added at time $s$ to the network evolves in this model according to

$$\frac{\partial d(s,t)}{\partial t} = (c_{+n} + 2c_{+e} + 2c_-)\frac{d(s,t)}{\int_0^t d(s,t)du},$$

a result that derives from a slight modification of an analytical model by Dorogovtsev and Mendes (9, 32). For sufficiently large time *t*, a power law distribution of degree, $P(d)=d^{-\gamma}$ with $\gamma>2$ emerges. (More specifically, $\gamma=2+c_{+n}/[c_{+n}+2c_{+e}-2c_{-}]$.) Thus, a simple local model including only empirically observed events suffices to explain the network's broad-tailed degree distribution. One important catch to any such model is that its results hold only in the limit of infinite time or infinite network size. However, all biological networks are small, fluctuate in size, and have evolved for a finite amount of time. It is thus best to study network evolution under the model's assumption numerically. Fig. 6 shows results of a simulation which starts out with a protein interaction network as observed (2), and shows the evolution of the degree distribution over 15,000 edge additions and deletions, a manifold turnover of the approximately 1000 interactions in the reported network. The ordinate axis shows the exponent $\gamma$ and its 95% confidence interval. The insets show the degree distribution after 0, 5000, and 10000 edge additions/deletions. Within the limits of statistical resolution, this distribution is invariant.

In sum, gene duplications do not alter network structure drastically. This is because duplicated protein-protein interactions diverge so rapidly and thoroughly that global network structure is left unchanged even after many gene duplications. Interaction turnover, on the other hand, is a more serious force. However, taken together, the observations that (i) the rate of interaction addition and deletion must be nearly balanced, (ii) interaction turnover affects preferentially highly connected proteins, and (iii) some added interactions add new proteins to the network, can explain the sustenance of the power law degree distribution. This analysis does not exclude the possibility of natural selection on the degree distribution as a whole but shows that it is not necessary.

**Caveats.** First, available protein interaction data is of limited quality. However, the pertinent global network structure is robust to variations in experimental technique. (Fig. 1). Second, considerable uncertainty is involved in estimating synonymous divergence of duplicate genes, especially for $K_s>1$. However, all divergence estimates are here used only to eliminate the most highly diverged genes, or to group duplicates into coarse age classes, never to base an argument on precise divergence dates. Third, while gene duplications may dominate genome evolution at short and intermediate time scales, exon or domain shuffling may have dominated early in the evolution of life. Because we have very little quantitative rate information about these latter processes, it is prudent to constrain the conclusions presented here to the intermediate time scale of several hundred Myrs for which data on evolutionary rates is available. Models that explicitly incorporate domain rearrangements may be more appropriate for larger time scales (33). Fourth, one can not say with certainty that divergence of interactions after gene duplications evolves only loss-of-interactions. However, circumstantial evidence suggests that degenerative mutations are rampant after gene duplications (27, 34). Such mutations would lead to an elevated rate at which interactions are lost. In addition, if most divergence after gene duplications were due to newly acquired interactions, the interaction density of the network would increase drastically over time. There is no indication for such an explosion of interaction density. This argument also speaks to the fifth caveat, which is that the rate at which interactions are added to the network might be coupled with gene duplications. It makes it seem less likely that recently duplicated genes acquire new interactions at increased rates. In addition, none of the observed cross-interactions between paralogues involve recently duplicated paralogues. Taken together, these observations suggest that the estimated rates of interaction turnover are also valid for extremely distant paralogous genes and for non-paralogous genes. Finally, there are uncertainties related to the small number of events on which some rate estimates are based. It is important to note that this limitation will not be overcome by improved data. It is mostly due to the limited number of gene duplicates in the (completely sequenced) yeast genome. I have thus taken care not to base any conclusions on a precise rate estimate and used rate information only qualitatively, such as to suggest that the rate of interaction turnover must be high enough to influence global network structure. A corollary to the limited resolution of the data is that the precise rates of node addition and interaction turnover may never be known. However, the ratios of these values used for the model are consistent with a key observation, that the rate of interaction addition must roughly balance that of interaction deletion, and that adding interactions often leads to adding proteins to the network. Although limited data resolution is likely to preclude any further statement, the observed processes are sufficient to sustain a power law distribution involving only local rules. They can explain network evolution without natural selection on global network structure.

**A question as old as biology**. Whether the *structure* of an organismal feature can provide information about some aspect of its *function* is a question as old as biology itself. It applies to every level of organization, from the arrangement of vertebrate bones to the conformation of proteins. It is key to philosophical debates central to biology, such as that between selectionists and neutralists. With the question's long history also come many cautionary tales. They range from Aristotle's infamous identification of the brain with a blood-cooling device, to the just-so-stories rampant in evolutionary biology.

Caution is thus necessary when postulating that natural selection on a global feature of a cellular network sustains this very feature. To be sure, a direct experimental test of this postulate for the power law degree distribution seems nearly impossible, as it requires generating a whole network with a different degree distribution and observing its performance in a living organism. However, indirect evidence can be obtained. Consider the example of metabolic networks. Abiotic chemical reaction networks, networks that have never been under the influence of natural selection, also show a power law degree distribution (35). This observation suggests that such a distribution may be inherent to any chemical reaction network. Consequently, it substantially weakens the selectionist case (4). The approach I took here was to explain network evolution from empirical observations and local rules without invoking natural selection on the degree distribution itself. It is, however, necessary to be aware that natural selection may be involved in many other ways. It may be involved in the addition and deletion of individual interactions, and thus act on a local scale. It may be responsible for the approximately balanced rate of interaction addition and deletion observed from the data. And it may have shaped the many other global features of this network. Identifying selection's role in shaping global network structures will doubtlessly provide a fruitful avenue towards identifying aspects of network function. But it is equally fraught with a danger that misled countless students of organisms all the way back to Aristotle.

**Acknowledgments**

Financial support through NIH grant GM063882-01 is gratefully acknowledged.

## Figure Captions

**Fig. 1.: The power-law degree distribution is a robust feature of the protein interaction network independent of experimental approach.** All three panels show a double logarithmic plot of the number of proteins (y-axis) with a given degree (x-axis), together with power law exponents and their 95% confidence intervals, as estimated by a linear regression analysis. Data for **(a)** and **(b)** are from two large-throughput two-hybrid experiments refs. 2,19 and ref. 1, respectively; **(c)** shows non-two hybrid data (28), as explained in detail below.

Data shown in (a) comprises 899 pairwise interactions among 985 yeast proteins, as reported in (2), and is available from http://depts.washington.edu/sfields/projects/YPLM/Nature-plain.html (obtained on February 15, 2000) as a list of pairwise interactions. I converted this list into a graph whose nodes represent proteins and whose edges correspond to protein interactions. The resulting protein interaction graph has $n$=985 proteins that engage in $k$=899 pairwise interactions. All reported graph analyses involve exhaustive enumeration using algorithms implemented in LEDA (36). The data for (b) stems from an independent high throughput experiment (1) also using the yeast two-hybrid assay. Its results are available from http://genome.c.kanazawa-u.ac.jp/Y2H. From these results I obtained in May 2001 a "core" data set of interactions confirmed in triplicate (1). The resulting protein interaction graph has $n$=780 proteins and $k$=747 interactions. To analyze protein interaction data not relying on the two-hybrid assay, I obtained information on physical interactions among yeast proteins from the MIPS database (28) at http://mips.gsf.de/proj/yeast/CYGD/db/index.html. I eliminated from this data all protein interactions confirmed by two-hybrid experiments. The remaining $k=899$ interactions involve $n=680$ proteins.

**Fig. 2.: Protein interactions and gene duplication. (a)** Mean and one standard deviation of the number of interactions per proteins with (left bar) and without (right bar)

paralogues in the yeast genome. All yeast duplicates, regardless of divergence were used in this analysis. **(b)** Circles stand for proteins, lines for interactions among proteins. After a gene duplication, the (identical) products P and $P^*$ of a duplicate gene interact with the same proteins. **(c)** The yeast protein interaction network contains many duplicate genes. Black circles represent proteins, black lines interactions between proteins, and red lines connect paralogous proteins. **(d)** Common interactions diverge rapidly after duplication. The x-axis corresponds to paralogous gene pairs in the protein interaction network binned according to the fraction of synonymous substitutions at synonymous sites, $K_s$. For each gene pair in each bin, I determined the number of shared interactions to identical third proteins, and divided it by the total number of interactions of the two proteins. The shown fraction of shared interactions is the average of this value over all gene pairs in a bin.

Data on yeast gene duplicates was kindly provided by John Conery (University of Oregon, Department of Computer Science) and was generated as described in (34). Briefly, gapped BLAST (37) was used for pairwise amino acid sequence comparisons of all yeast open reading frames as obtained from GenBank. All protein pairs with a BLAST alignment score greater than $10^{-2}$ were retained for further analysis. Then, the following conservative approach was followed to retain only unambiguously aligned sequences. Using the protein alignment generated by BLAST as a guide, a sequence pair was scanned to the right of each alignment gap. All sequence from the end of the gap through the first "anchor" pair of matched amino acids was discarded. All subsequent sequence (exclusive the anchor pair of amino acids) was retained if a second pair of matching amino acids was found within less than six amino acids from the first. This procedure was then repeated to the left of each alignment gap (see ref. 34 for more detailed description and justification.). The retained portion of each amino acid sequence alignment was then used jointly with DNA sequence information to generate nucleotide sequence alignments of genes. For each gene pair in this data set, the fraction $K_s$ of synonymous (silent) substitutions per silent site, as well as the fraction $K_a$ of replacement substitutions per replacement site were estimated using the method of Li (38). For the analysis in Figure 2a all paralogous genes were used, for the analysis in Figure 2d only paralogues with $K_s<1.25$.

**Fig. 3.: Duplication and divergence, taken together, do not affect degree distribution.** Both panels show a double logarithmic plot of the degree distribution in the protein network as reported by ref. 2, and in the same network after 1000 and 2000 gene duplications. With an effective rate of $8.3 \times 10^{-4}$/gene/Myr, about 1000 gene duplications are expected in the network every 1.2 Gyr. The numbers entitled "exp" are the exponents $\gamma$ of the degree distribution, $P(d) \propto d^{-\gamma}$. The 95% confidence intervals of these exponents are greater than 0.35 in all six cases shown. Thus, the distributions are statistically distinguishable.

The results stem from a numerical simulation of network evolution, where I repeatedly duplicated individual genes chosen at random from the network, and let interactions diverge as follows after each duplication. **(a)** Asymmetric divergence. Functional divergence between gene duplicates generally occurs asymmetrically, i.e., one gene product retains more molecular interactions than the other. This is the divergence pattern observed empirically (39). To emulate this scenario, I chose one of the two duplicate genes at random, and eliminated each interaction of this gene independently with probability $1\text{-}f\text{=}0.85$. This procedure ensures not only that divergence is asymmetric, but also that an average of 15% of shared interactions remain after the duplication, a value close to the maximum of that observed in the empirical data (see text). **(b)** Symmetric divergence. I chose and eliminated one of the two interactions in each redundant interaction pair of two gene duplicates (Fig. 2b) with probability ½. After this procedure, the expected number of retained interactions per gene is half the number before duplication. In this sense, divergence is symmetric. I eliminated proteins without remaining interactions after this procedure.

**Fig. 4: Evolution of new interactions. (a)** Interactions between duplicate genes may evolve along two different routes. First, a gene product may have been a self-interactor before duplication. In this case, observed self-interactions and interactions between duplicates are a remnant of self-interaction before duplication. Second, the interactions may have evolved *de novo* after the duplication. **(b)** Number of paralogous gene pairs observed in the yeast protein interaction networks with the indicated combination of self-

and cross-interaction. The left and middle columns represent data from refs. 2 and 1, respectively. The right-most column represents non two-hybrid data (28). Notice the abundance of duplicates without self-interactions and the small number of gene pairs where both genes are self-interacting. The total number of nodes in each of these networks, including nodes that show only self-interactions is (from left to right) 999, 971, and 680. Within each of the five classes of gene pairs shown, multiple pairs belonging to the same gene family were eliminated before analysis. Each gene family is thus represented here by only one gene pair, in order to eliminate statistical bias due to large gene families, in contrast to an earlier analysis based on a smaller data set (10).

**Fig. 5: Preferential attachment in protein interaction networks.** The abscissa shows the degree $d$ of the protein. The ordinate axis shows the likelihood $P_d$ that a protein of degree $d$ has evolved new interactions. To obtain $P_d$ for each $d$, I considered all paralogous gene pairs in the data pooled from refs. 1,2, and 24 with (i) $K_s<3$, (ii) cross-interactions between the paralogues, and (iii) no self-interactions. To avoid statistical bias, only one pair of genes from each multigene family is included in the analysis. Among these paralogues, I determined the number $I_d$ of those proteins that had $d$ interactions to proteins different from its paralogous partner. To account for the fact that proteins of different degree occur at different frequencies in the network, I then divided this number by the relative frequency $f_d$ of proteins of degree $d$ in the network, and normalized the resulting quantity to obtain $P_{d,}$ i.e., $P_d=(I_d/f_d))/\Sigma_d\,(I_d/f_d)$. There is a strong, approximately linear association between protein degree and the likelihood to evolve new interactions.

**Fig. 6: Power law degree distribution through local rules.** Shown are results from a stochastic simulation of network evolution, beginning with the network as reported by ref. 2. The ordinate axis shows the power law exponent $\gamma$ and its 95% confidence interval, as obtained from a linear regression analysis. The insets show the complete degree distribution after 0, 5000, and 10000 edge additions/deletions. Within the limits of statistical resolution, this distribution is invariant over the 15,000 added and deleted

interactions shown on the abscissa (50-140 Myrs of evolution). Because the network as reported contains less than 1000 interactions, a turnover of 15,000 interactions means that each and every interaction is turned over many times. At each time step shown, an interaction was added between network proteins with probability $c_{+e}=0.3$, a protein was added to the network (via one interaction to a network protein) with probability $c_{+n}=0.7$, and an interaction was eliminated with probability $c_{-}=1$. These parameter values imply that the rate at which interactions are added and eliminated is approximately equal, and that a fraction of new edge additions also involves the addition of new nodes, as observed empirically. Beyond these requirements, the empirical data does not provide sufficient resolution to estimate these relative rates precisely. They were thus chosen such that the overall number of proteins remains roughly constant over the time interval shown.

To ensure that network evolution follows the rule of preferential attachment I followed the following procedure. To add an interaction between network proteins, I first determined the sum $s$ of all degrees of network proteins. Two non-adjacent network proteins $u$ and $v$ were then chosen at random. I then chose a random number $r$ uniformly distributed on the interval (0,1). If $r<d(v)/s$, where $d(v)$ is the degree of $v$ then I established an interaction between $u$ and $v$. If not, I repeated the process of chosing $v$ and generating $r$ until $r<d(v)/s$ and a new interaction could be established. This ensures that newly added edges connect preferentially to highly connected nodes. To add a new node to the network, I followed an identical procedure, except that I did not choose the node $u$ from within the network, but generated it as an isolated node. Finally, to eliminate interactions, I simply chose one interaction at random and eliminated it. If this resulted in a node to be isolated, I eliminated this node as well. Because edges are more likely to be attached to highly connected nodes, this ensures that interactions are preferentially eliminated from highly connected nodes. The regression analysis in the plot was carried out only if none of the frequencies of proteins with $1<d<5$ was zero. In all other cases, data is shown as missing in the plot.

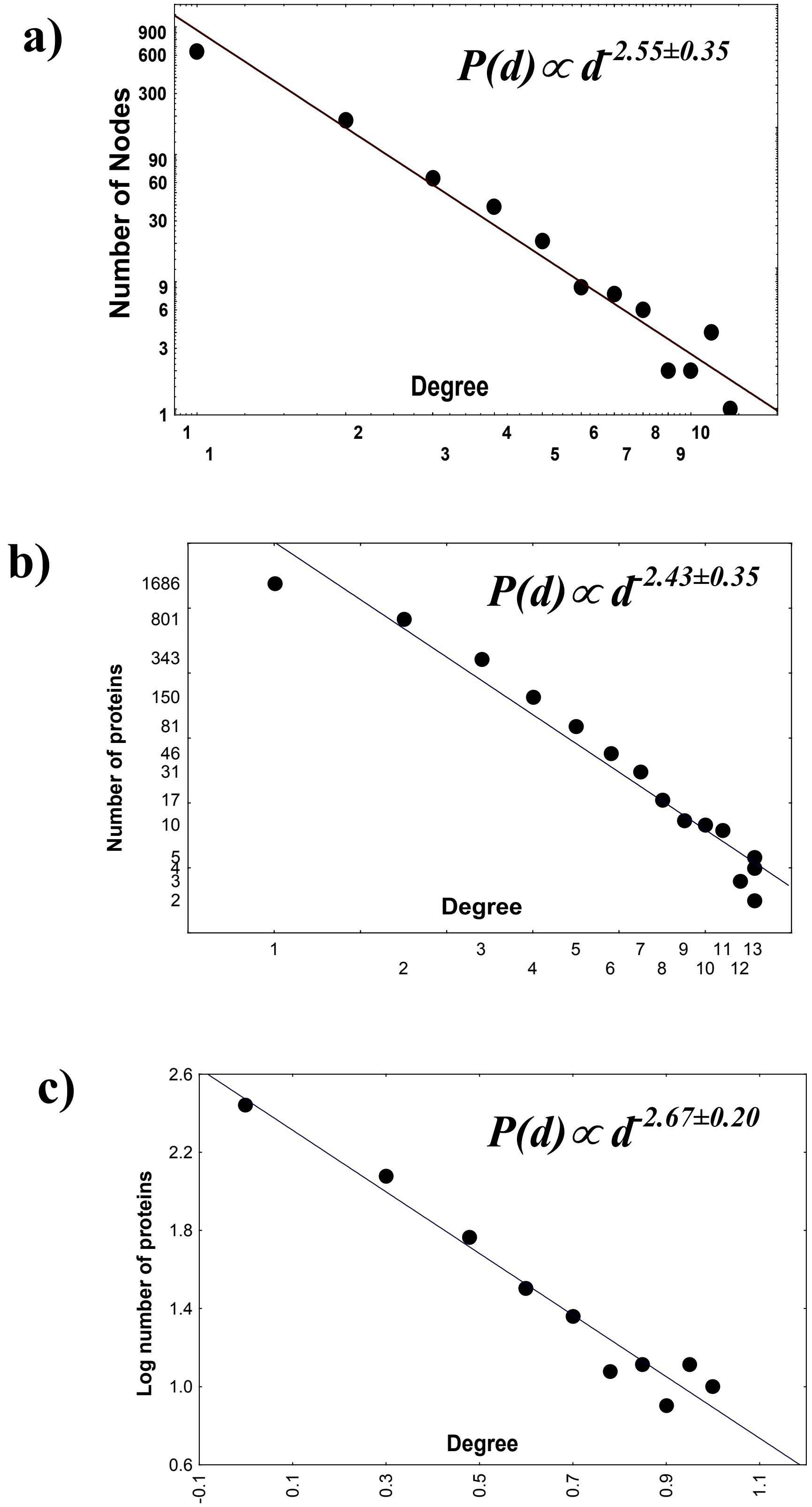

Fig. 1

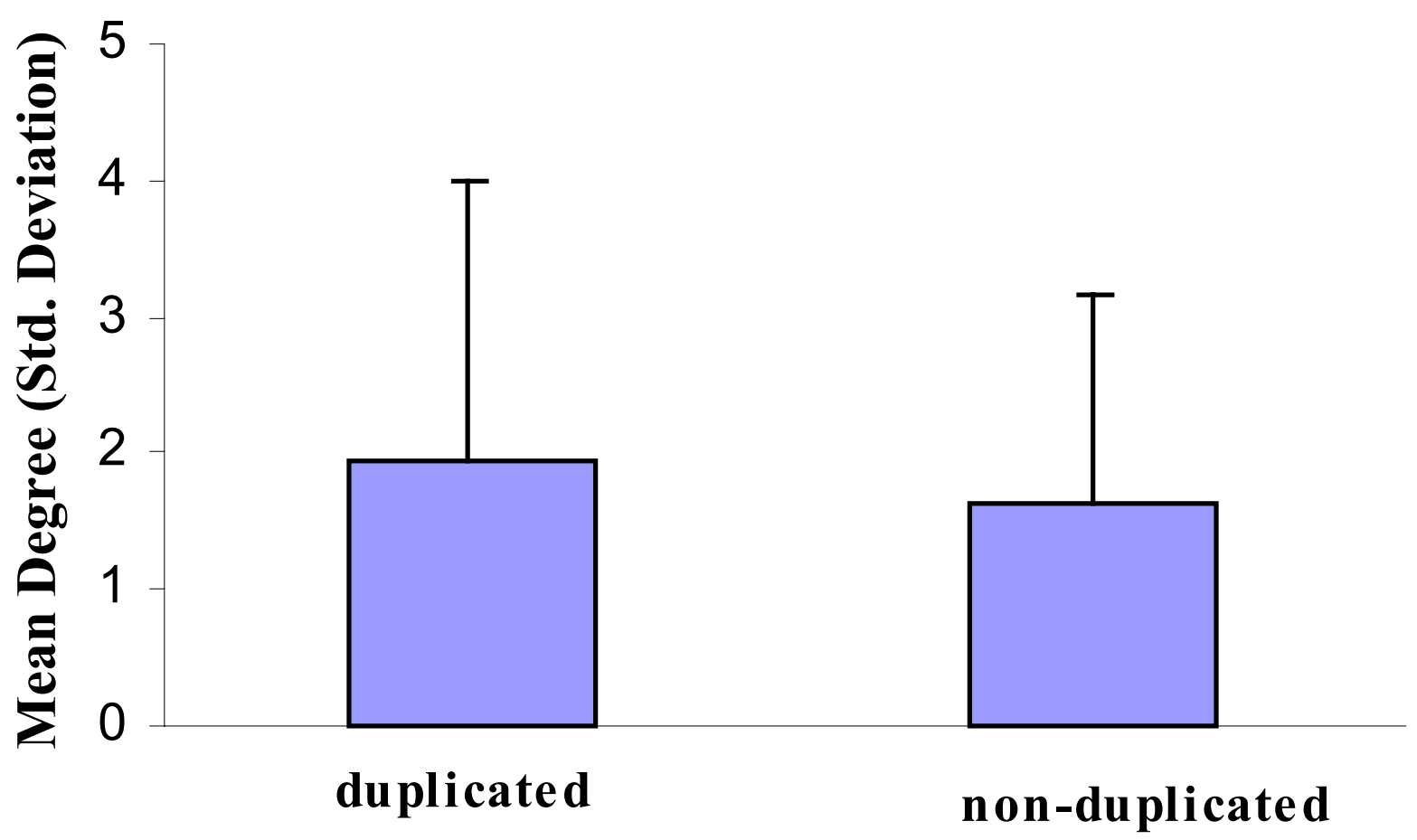

**Fig. 2a**

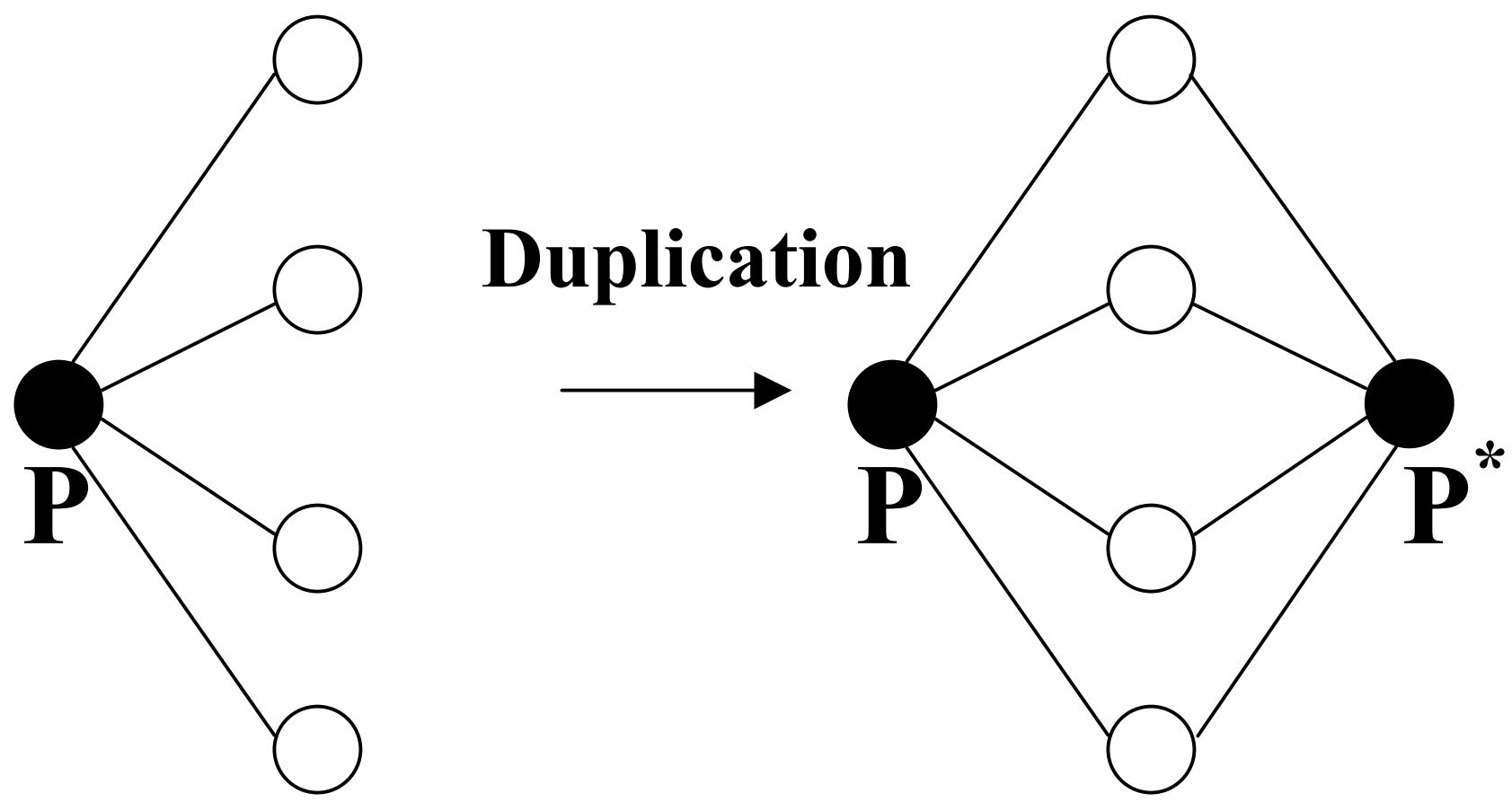

Fig. 2b

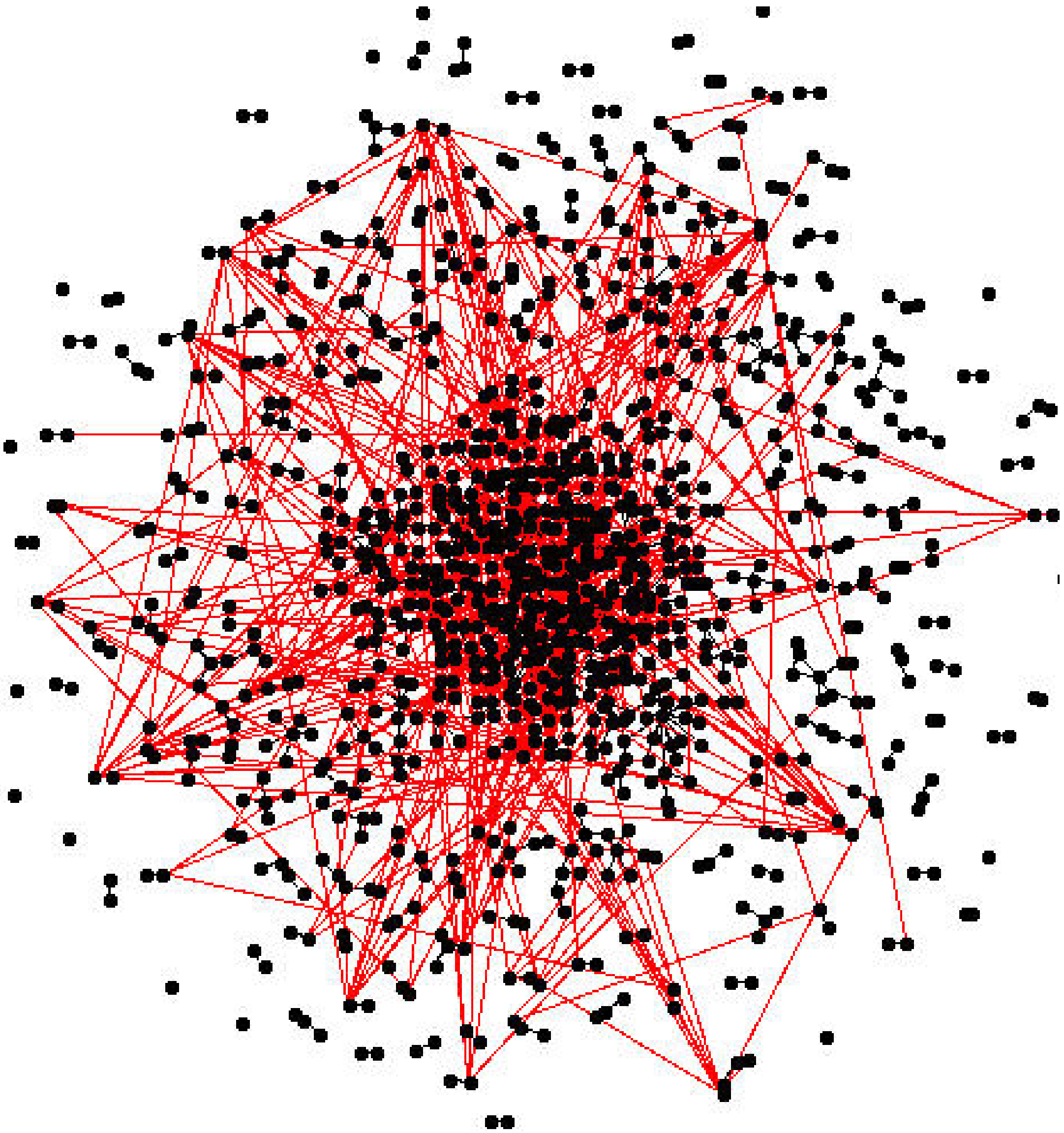

**Fig. 2c**

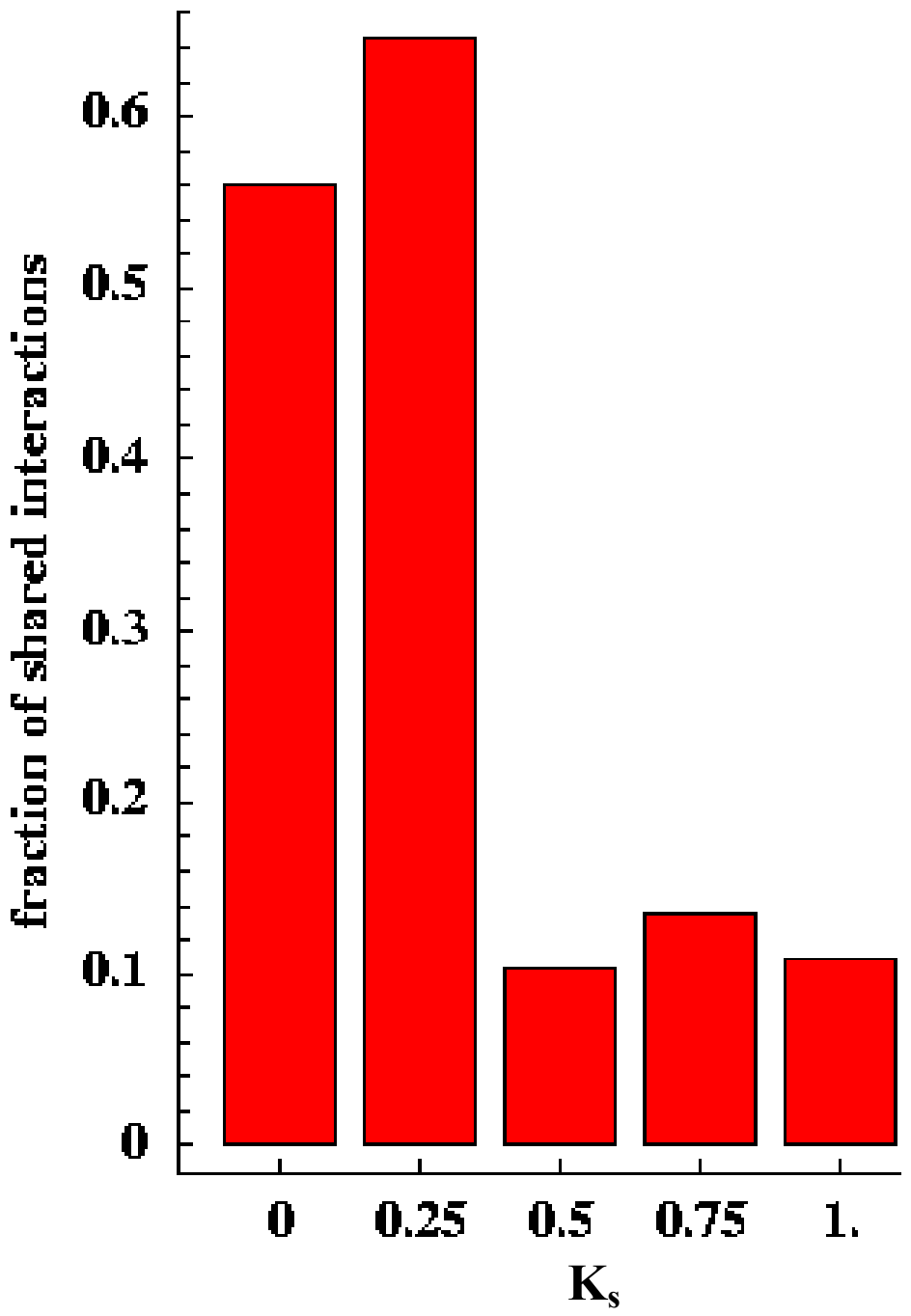

**Fig. 2d**

**a**

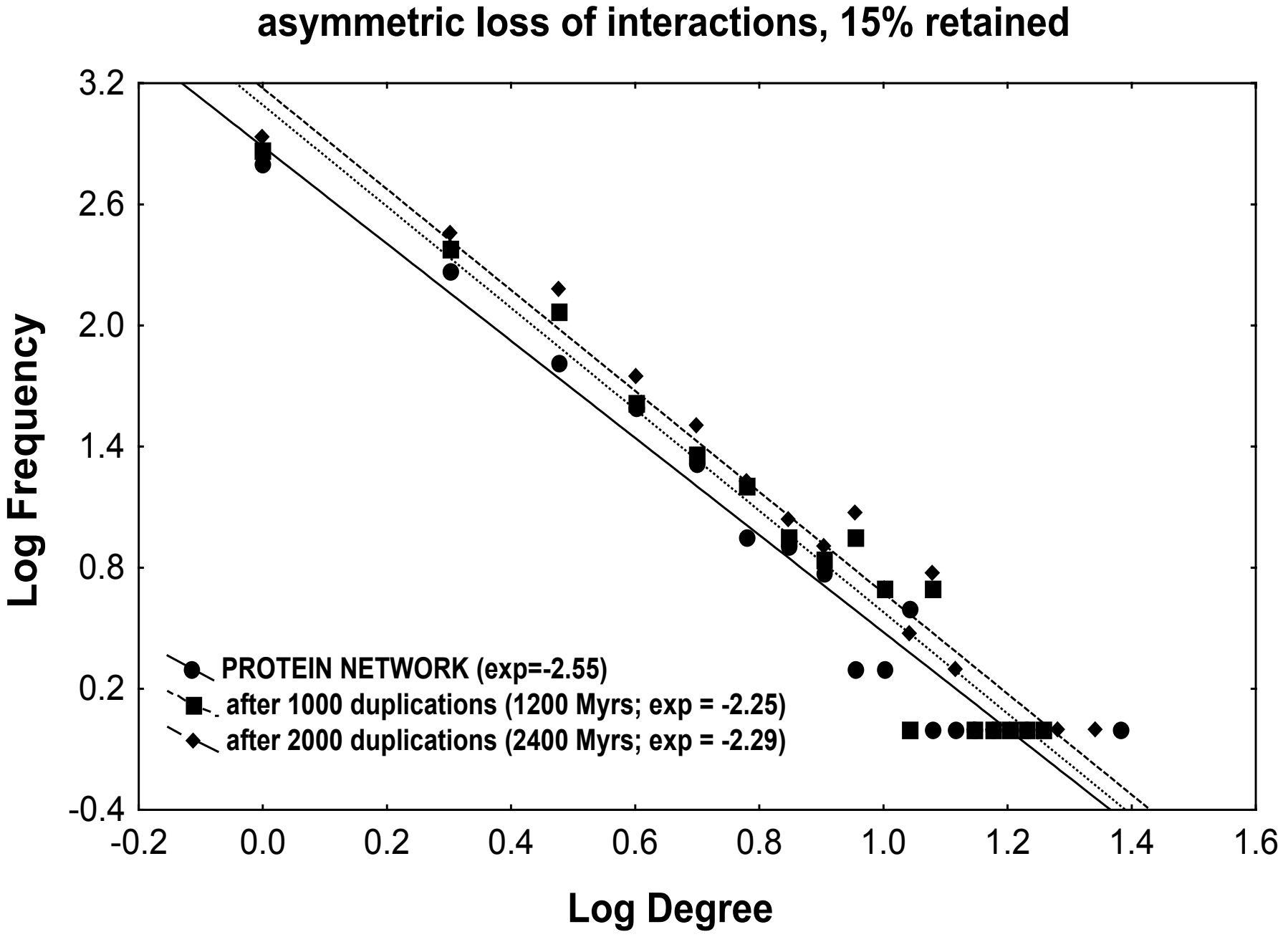
asymmetric loss of interactions, 15% retained
Log Frequency
Log Degree
PROTEIN NETWORK (exp=-2.55)
after 1000 duplications (1200 Myrs; exp = -2.25)
after 2000 duplications (2400 Myrs; exp = -2.29)
3.2
2.6
2.0
1.4
0.8
0.2
-0.4
-0.2
0.0
0.2
0.4
0.6
0.8
1.0
1.2
1.4
1.6

**b**

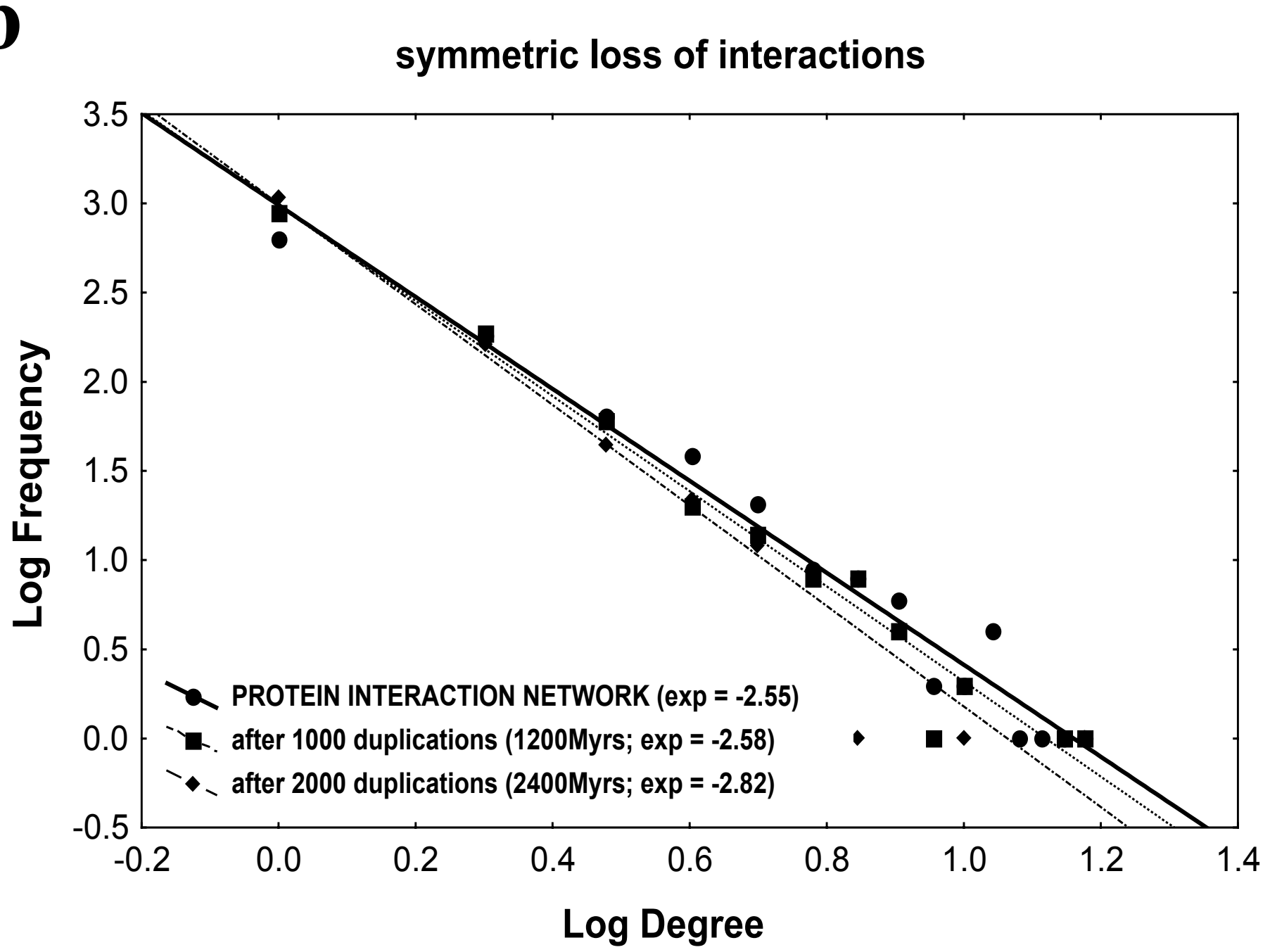

**Fig. 3**

a)

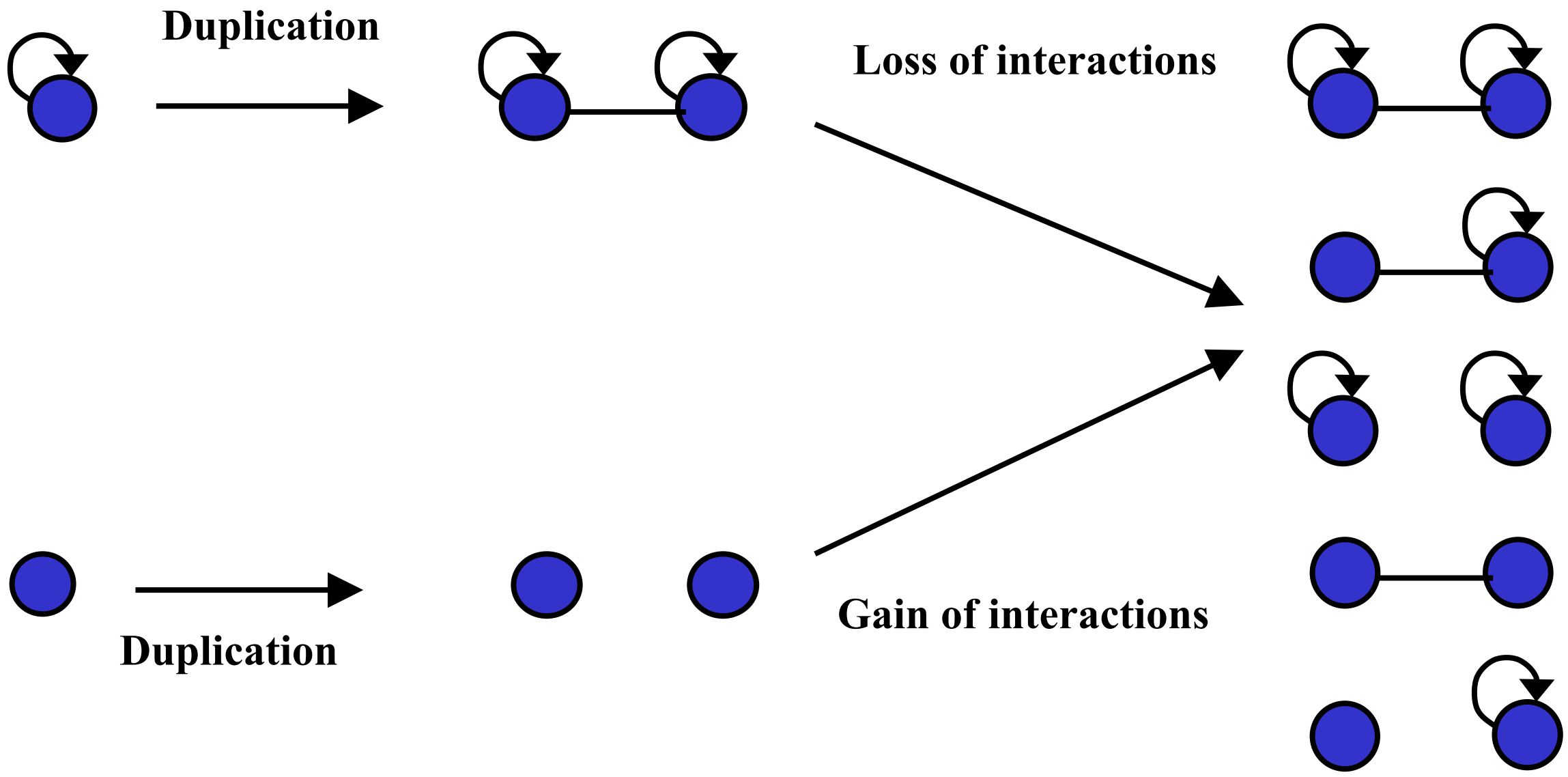

b)

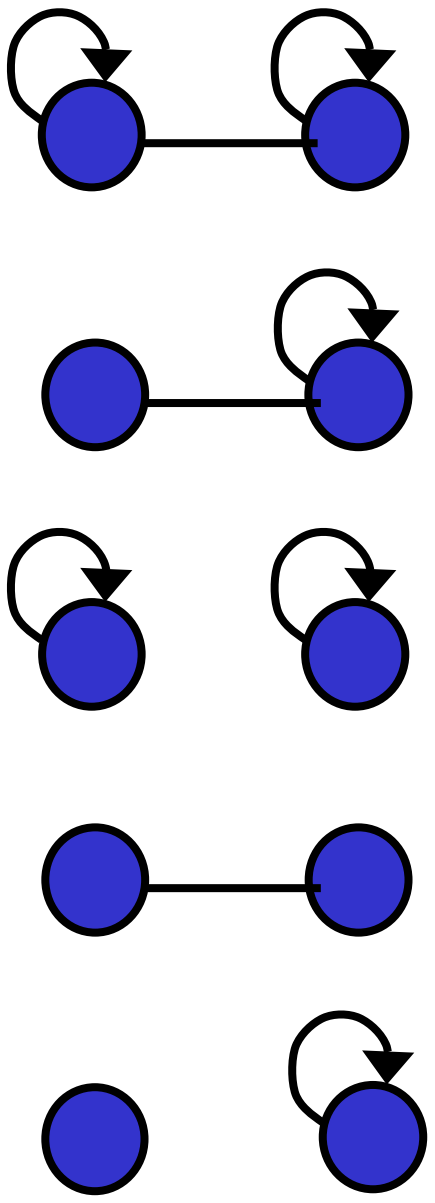

| two-hybrid | | non two-hybrid |
|---|---|---|
| 1 | 1 | 0 |
| 4 | 2 | 1 |
| 0 | 2 | 1 |
| 10 | 13 | 11 |
| 4 | 4 | 9 |

**Fig. 4**

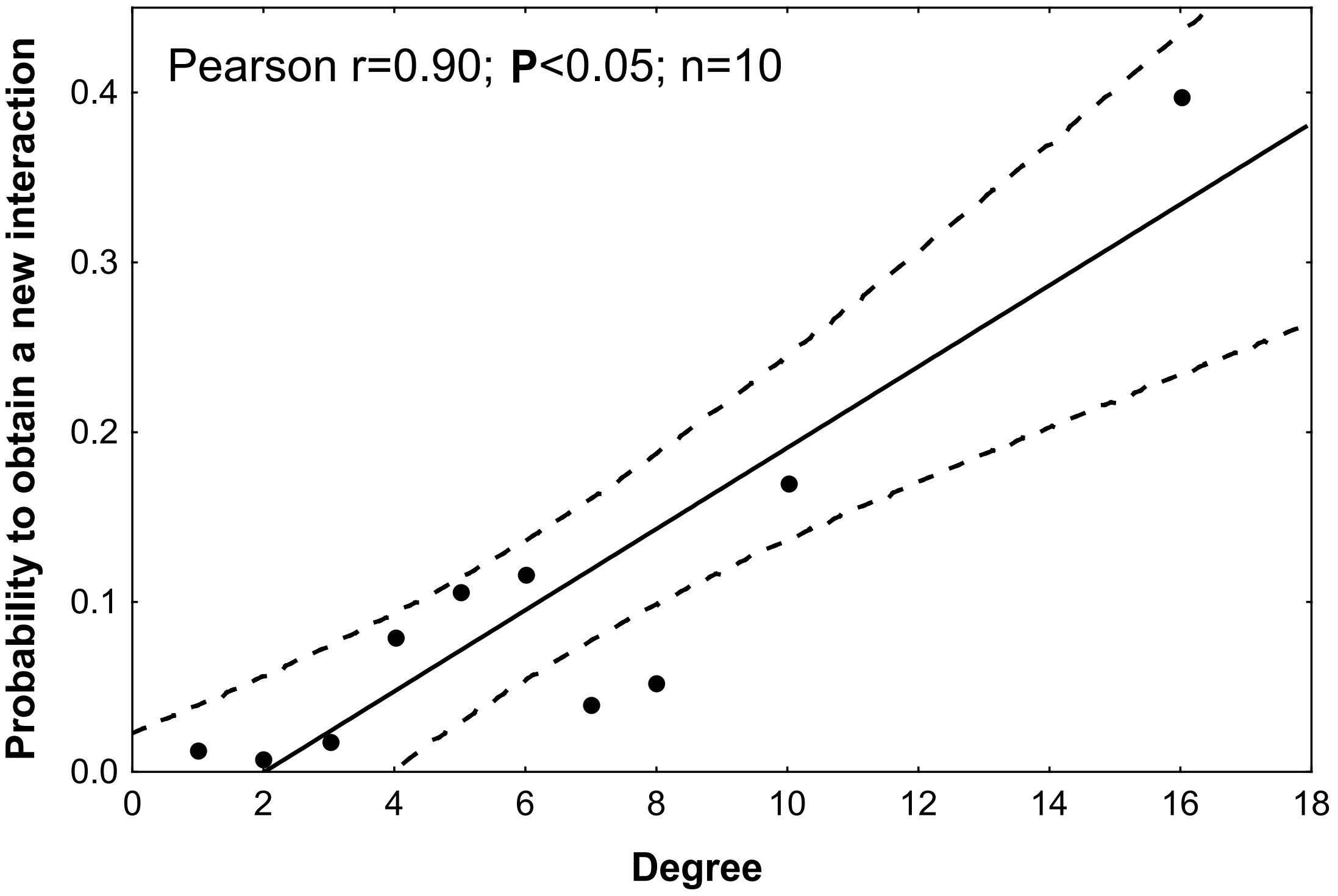

**Fig. 5**

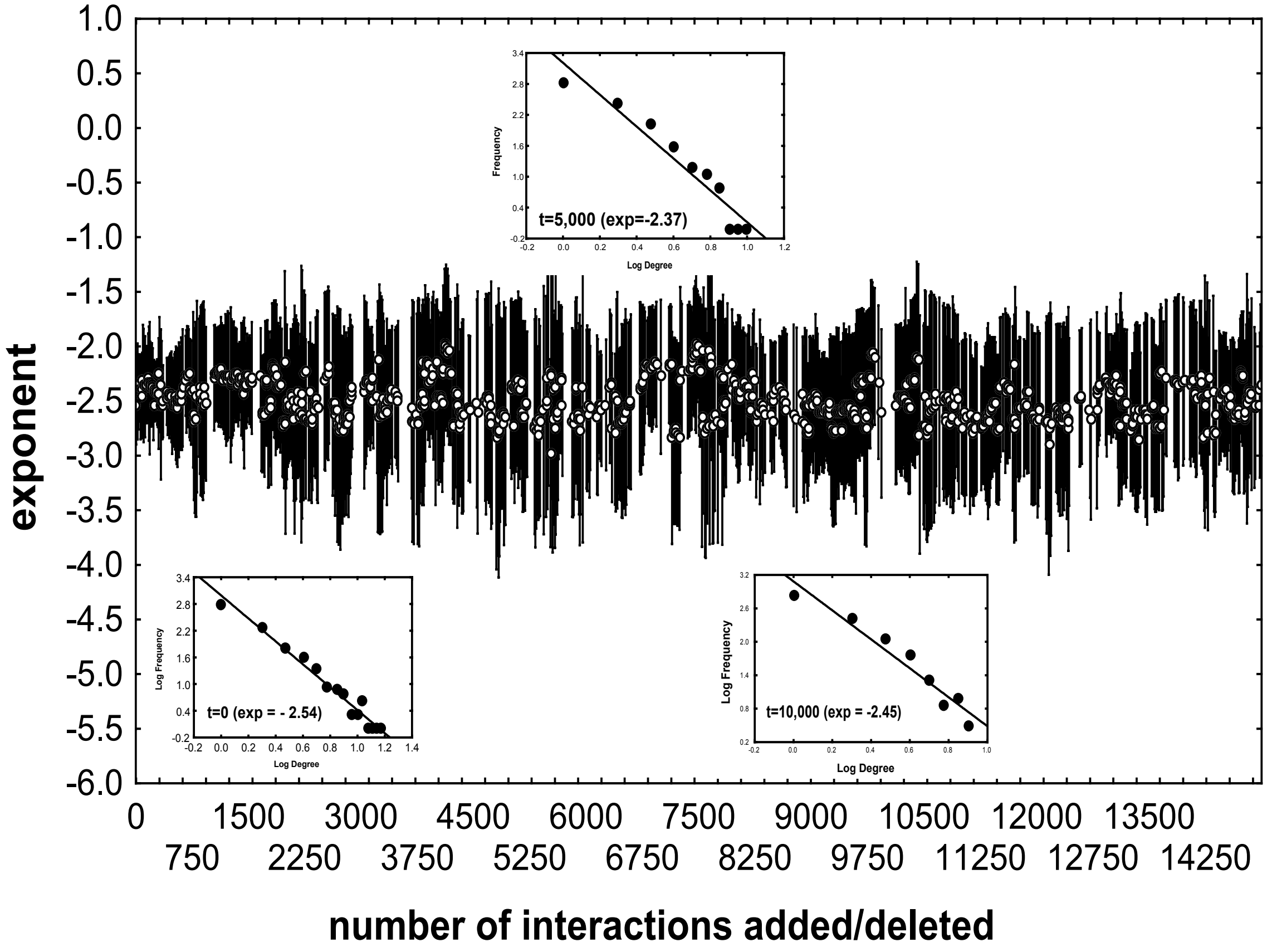

**Fig. 6**